\documentclass[10pt, conference]{IEEEtran}

\usepackage{lmodern}

\usepackage{url}
\usepackage{hyperref} 

\usepackage{amsmath,amssymb,wasysym,mathtools,esvect,dsfont}

\usepackage{graphicx}
\usepackage{xcolor}

\usepackage{float}

\usepackage{tikz}
\usetikzlibrary{automata}
\usetikzlibrary{arrows}
\usetikzlibrary{shadows}
\usetikzlibrary{decorations.text}
\usetikzlibrary{circuits.logic.US} 
\usetikzlibrary{matrix}

\usepackage{subcaption}

\usepackage{tabularx}
\newcolumntype{L}[1]{>{\raggedright\arraybackslash}p{#1}} 
\newcolumntype{C}[1]{>{\centering\arraybackslash}p{#1}} 
\newcolumntype{R}[1]{>{\raggedleft\arraybackslash}p{#1}} 
\usepackage{multirow}
\usepackage{adjustbox}
\usepackage{booktabs}
\usepackage{colortbl}
\usepackage{listings}
\usepackage{algorithm}
\usepackage[noend]{algpseudocode}
\algrenewcommand\algorithmicrequire{\textbf{\ \ Input:}}
\algrenewcommand\algorithmicensure{\textbf{Output:}}

\usepackage{xspace}
\usepackage{pifont}
\usepackage{comment}

\usepackage[nolist]{acronym}

\begin{acronym}

\acro{3DES}{Triple-DES}

\acro{AES}{Advanced Encryption Standard}
\acro{ALU}{Arithmetic Logic Unit}
\acro{API}{Application Programming Interface}
\acro{ARX}{Addition Rotation XOR}
\acro{ATPG}{Automatic Test Pattern Generation}
\acro{ASIC}{Application Specific Integrated Circuit}
\acro{ASIP}{Application Specific Instruction-Set Processor}
\acro{AS}{Active Serial}

\acro{BDD}{Binary Decision Diagram}
\acro{BGL}{Boost Graph Library}
\acro{BNF}{Backus-Naur Form}
\acro{BRAM}{Block-Ram}

\acro{CBC}{Cipher Block Chaining}
\acro{CFB}{Cipher Feedback Mode}
\acro{CFG}{Control Flow Graph}
\acro{CLB}{Configurable Logic Block}
\acro{CLI}{Command Line Interface}
\acro{COFF}{Common Object File Format}
\acro{CPA}{Correlation Power Analysis}
\acro{CPU}{Central Processing Unit}
\acro{CRC}{Cyclic Redundancy Check}
\acro{CTR}{Counter}

\acro{DC}{Direct Current}
\acro{DES}{Data Encryption Standard}
\acro{DFA}{Differential Fault Analysis}
\acro{DFT}{Discrete Fourier Transform}
\acro{DIP}{Distinguishing Input Pattern}
\acro{DLL}{Dynamic Link Library}
\acro{DMA}{Direct Memory Access}
\acro{DNF}{Disjunctive Normal Form}
\acro{DPA}{Differential Power Analysis}
\acro{DSO}{Digital Storage Oscilloscope}
\acro{DSP}{Digital Signal Processing}
\acro{DUT}{Design Under Test}

\acro{ECB}{Electronic Code Book}
\acro{ECC}{Elliptic Curve Cryptography}
\acro{EEPROM}{Electrically Erasable Programmable Read-only Memory}
\acro{EMA}{Electromagnetic Emanation}
\acro{EM}{electro-magnetic}

\acro{FFT}{Fast Fourier Transformation}
\acro{FF}{Flip Flop}
\acro{FI}{Fault Injection}
\acro{FIB}{Focused Ion Beam}
\acro{FIR}{Finite Impulse Response}
\acro{FPGA}{Field Programmable Gate Array}
\acro{FSM}{Finite State Machine}

\acro{GUI}{Graphical User Interface}

\acro{HDL}{Hardware Description Language}
\acro{HD}{Hamming Distance}
\acro{HF}{High Frequency}
\acro{HSM}{Hardware Security Module}
\acro{HW}{Hamming Weight}

\acro{IC}{Integrated Circuit}
\acro{IO}[I/O]{Input/Output}
\acro{IOB}{Input Output Block}
\acro{IOT}[IoT]{Internet of Things}
\acro{IP}{Intellectual Property}
\acro{ISA}{Instruction Set Architecture}
\acro{IV}{Initialization Vector}

\acro{JTAG}{Joint Test Action Group}

\acro{KAT}{Known Answer Test}

\acro{LFSR}{Linear Feedback Shift Register}
\acro{LSB}{Least Significant Bit}
\acro{LUT}{Look-up table}

\acro{MAC}{Message Authentication Code}
\acro{MLE}{Maximum Likelihood Estimation}
\acro{MIPS}{Microprocessor without Interlocked Pipeline Stages}
\acro{MMIO}{Memory Mapped \acl{IO}}
\acro{MSB}{Most Significant Bit}

\acro{NASA}{National Aeronautics and Space Administration}
\acro{NSA}{National Security Agency}
\acro{NVM}{Non-Volatile Memory}

\acro{OFB}{Output Feedback Mode}
\acro{OISC}{One Instruction Set Computer}
\acro{ORAM}{Oblivious Random Access Memory}
\acro{OS}{Operating System}

\acro{PAR}{Place-and-Route}
\acro{PCB}{Printed Circuit Board}
\acro{PC}{Personal Computer}
\acro{PFA}{Persistent Fault Analysis}
\acro{PRNG}{Pseudo Random Number Generator}
\acro{PS}{Passive Serial}
\acro{PUF}{Physically Unclonable Function}

\acro{RISC}{Reduced Instruction Set Computer}
\acro{RNG}{Random Number Generator}
\acro{ROBDD}{reduced ordered binary decision diagram}
\acro{ROM}{Read-Only Memory}
\acro{ROP}{Return-oriented Programming}
\acro{RTL}{Register Transfer Level}

\acro{SAT}{Boolean satisfiability}
\acro{SBox}{Substitution Box}
\acro{SEI}{Squared Euclidean Imbalance}
\acro{SCA}{Side-Channel Analysis}
\acro{SFA}{Statistical Fault Analysis}
\acro{SFLL}{Stripped-Functionality Logic Locking}
\acro{SHA}{Secure Hash Algorithm}
\acro{SNR}{Signal-to-Noise Ratio}
\acro{SPA}{Simple Power Analysis}
\acro{SPFA}{Statistical Persistent Fault Analysis}
\acro{SPI}{Serial Peripheral Interface Bus}
\acro{SPN}{Substitution-Permutation Network}
\acro{SRAM}{Static Random Access Memory}

\acro{TRNG}{True Random Number Generator}

\acro{UART}{Universal Asynchronous Receiver Transmitter}
\acro{UHF}{Ultra-High Frequency}

\acro{WISC}{Writeable Instruction Set Computer}

\acro{XDL}{Xilinx Description Language}
\acro{XTS}{XEX-based Tweaked-codebook with ciphertext Stealing}

\end{acronym}

\makeatletter
\newcommand{\removeifnextchar}[2]{%
    \begingroup
    \ltx@LocToksA{\endgroup#2}%
    \ltx@ifnextchar@nospace{#1}{%
        \def\next{\the\ltx@LocToksA}%
        \afterassignment\next
        \let\scratch= %
    }{%
        \the\ltx@LocToksA
    }%
}

\newcommand{\ie}{\protect\removeifnextchar{,}{i.e.,\xspace}}
\newcommand{\eg}{\protect\removeifnextchar{,}{e.g.,\xspace}}
\newcommand{\etal}{\protect\removeifnextchar{.}{et~al.\@ifnextchar{.}{}{\@ifnextchar{,}{}{~}}}}

\makeatother

\newcommand{\revise}[2]{#2}

\newcommand{\romannumber}[1]{\uppercase\expandafter{\romannumeral#1}}
\usepackage{todonotes}

\ifdefined\algorithmautorefname
\else
\newcommand{\algorithmautorefname}{Algorithm}
\fi 

\ifdefined\definitionautorefname
\else 
\newcommand{\definitionautorefname}{Definition}
\fi

    \ifdefined\algorithmautorefname
    \renewcommand{\algorithmautorefname}{Algorithm}
    \fi 
    \ifdefined\definitionautorefname
    \renewcommand{\definitionautorefname}{Definition}
    \fi

\newcommand{\Figure}[1]{\autoref{#1}}

\newcommand{\Section}[1]{\autoref{#1}}
\newcommand{\Table}[1]{\autoref{#1}}

\usetikzlibrary{shapes.geometric}
\usetikzlibrary{patterns}

\title{SPFA: SFA on Multiple Persistent Faults}

\author{\IEEEauthorblockN{Susanne Engels}
	\IEEEauthorblockA{Horst G\"ortz Institute for IT-Security\\
		Ruhr University Bochum\\
		Bochum, Germany\\
		Email: susanne.engels@rub.de}
	\and
	\IEEEauthorblockN{Falk Schellenberg, Christof Paar\IEEEmembership{Fellow,~IEEE}}
	\IEEEauthorblockA{Max Planck Institute for \\
		Cybersecurity and Privacy\\
		Bochum, Germany\\
		Email: firstname.lastname@csp.mpg.de}
}

\begin{document}

\maketitle

\acused{AES}

\begin{abstract}
	For classical fault analysis, a transient fault is required to be injected during runtime, e.g., only at a specific round.
	Instead, Persistent Fault Analysis (PFA) introduces a powerful class of fault attacks that allows for a fault to be present throughout the whole execution.
    One limitation of original PFA as introduced by Zhang et al.~at CHES'18 is that the faulty values need to be known to the adversary.
 	While this was addressed at a follow-up work at CHES'20, the solution is only applicable to a single faulty value.
	\revise{}{Instead, we use the potency of Statistical Fault Analysis (SFA) in the persistent fault setting, presenting \ac{SPFA} as a more general approach of PFA. }
 	As a result, any or even a multitude of unknown faults that cause an exploitable bias in the targeted round can be used to recover the cipher's secret key.
 	Indeed, the undesired faults in the other rounds that occur due the persistent nature of the attack converge to a uniform distribution as required by SFA.
    We verify the \revise{}{effectiveness} of our attack against LED and AES. 
\end{abstract}


\section{Introduction}
\label{pf::sec::intro}

In 1997, Boneh \etal were the first to present that errors during the computation of RSA with CRT can be exploited to reveal a prime factor of the public modulus N \cite{boneh:1997:eurocrypt}.
Their findings triggered further research on how to break cryptographic schemes by injecting a fault during its execution. 
The most prominent fault analysis technique is \acf{DFA} which exploits the difference between correct and faulty results \cite{biham:1997:crypto}.
With \ac{DFA}, it is possible to attack general \ac{SPN} structures used in block ciphers such as AES \cite{piret:2003:ches}.
Along with the differential approach, statistical approaches to exploit faults have been presented~\cite{dobraunig:2016:asiacrypt}.
If the adversary does not have precise knowledge of the fault model, \acl{SFA} featuring the \ac{SEI} can be used \cite{rivain:2009:ches}.
In contrast to \ac{DFA}, \ac{SFA} allows for faulty-ciphertexts-only attacks \cite{fuhr:2013:ftdc}.

Although various further analysis techniques against different cryptosystems have been proposed, almost all of them have in common that they need for the injected fault to be \textit{transient}.  
Transient faults only exist for a limited period of time after which the original value is restored. 
Opposed to this, a permanent fault can be regarded as a destructive manipulation of the device, which cannot be reverted.
For \textit{persistent} faults as third type, the faulty value is stored until it is overwritten or the device is reset.
For example, when a value within the memory storing an SBox is faulted, this means that the faulty value might be accessed at multiple rounds.
A successful attack in this scenario was presented by Zhang \etal in~\cite{zhang:2018:ches} as \ac{PFA}.
Unfortunately, the exact position (or value) of the fault had to be known \revise{}{or brute-forced}.
This limitation was improved recently by extracting the actual faulty value beforehand~\cite{zhang:2020:ches}.
However, this improvement can cope only with single-value faults.

We generalize those attacks on persistent faults and present an attack that can handle a multitude of unknown faulty values, e.g., in the SBox memory.
To this end, we extend the well-known concept of SFA by introducing persistent faults as used in PFA, resulting in \acf{SPFA}. 
Inherited from classical \ac{SFA}, we only require that the persistent fault creates some form of bias of internal states (which is usually the case, cf. \cite{DBLP:conf/fdtc/GhalatyYTS14,dobraunig:2016:asiacrypt}).
For \ac{SFA}, this bias as deviation from an uniform distribution is only visible for the correct key hypothesis.
Our basic assumption in the persistent setting is that faults occurring in the other rounds will only have a limited effect on the targeted round.
This is due to the still existing diffusion properties of the cryptographic primitive that will eventually spread the undesired faults to become closer to a uniform distribution.
Indeed, we show that our attack succeeds even if the fault is present throughout the whole encryption and allow for single and especially numerous affected bytes, e.g., within one S-Box.
To practically evaluate the potency of our attack, we perform \ac{SPFA} on two block ciphers, LED and AES.






\section{Background and Related Work}
\label{pf::sec::rel_work}

\enlargethispage{\baselineskip}

The permanency of a physical fault is often considered as one central parameter for fault attacks~\cite{DBLP:conf/cardis/GiraudT04,DBLP:conf/fdtc/VerbauwhedeKS11}.
A physical fault injection might result in a fault that is either:
\begin{itemize}
	\item transient, for example only for the duration of the physical effect, 
	\item persistent, e.g., when stored in memory or registers until the value is overwritten or the device is reset,
	\item permanent, sometimes also termed as destructive~\cite{DBLP:journals/tvlsi/KaraklajicSV13}.
\end{itemize}
Despite some early work on permanent faults targeting asymmetric schemes~\cite{DBLP:conf/acisp/YenMH03}, persistent or permanent fault analysis received little attention for a long time.

In contrast, there are numerous examples of physical faulting methods that indeed actually result in a non-transient behavior:
For example, \cite{DBLP:conf/fdtc/VerbauwhedeKS11} points to physical faults that might permanently destroy parts of the circuit, e.g., burnout or latch-up faults.
Further, Rowhammer-style attacks create persistent bit flips in DRAM~\cite{DBLP:conf/isca/KimDKFLLWLM14}.
The same holds for other storage elements, e.g., when focusing a laser beam on SRAM cells, a faulty value might be stored until the cell is overwritten~\cite{DBLP:conf/ches/SkorobogatovA02}.
The initial work  by Skorobogatov \etal triggered an arms race at which technology node one could still inject single bit faults into SRAM using lasers.
The established limit seems to be at 45nm feature size~\cite{DBLP:conf/cardis/SelmkeBHS15} where multiple cells become affected.
Indeed, more recent work targets much larger flip-flops instead of SRAM at 28nm~\cite{DBLP:conf/fdtc/DutertreBCCFFGH18}.
Swierczynski \etal observed control flow alterations when tempering with the bitstream configuration of FPGAs \cite{pawel:2018:tc}, which can be seen as a persistent fault attack as well.
Thus, especially for the latter methods, we might observe multiple persistent faults per fault injection.

The topic of persistent fault analysis was picked up in 2018 by Zhang \etal, introducing the notion of \ac{PFA} \cite{zhang:2018:ches}.
In 2019, Gruber \etal presented how to attack the CAESAR finalists COLM, Deoxys and OCB using \ac{PFA}, and confirmed their results using simulation \cite{gruber:2019:ftdc}.
Using the same strategies presented by Zhang \etal, the authors experimentally prove the applicability of \ac{PFA} and give numbers on the amount of faulty ciphertexts needed.
%
Caforio \etal use \ac{PFA} to recover secret Sbox tables using the example of PRESENT and AES \cite{caforio:2019:space} while Pan \etal investigate the security of masking schemes against PFA, practically showing that one persistent fault is enough to break masking at any masking order $d$ \cite{pan:2019:date}.
The initial work by Zhang \etal was later improved in \cite{zhang:2020:ches}, reducing the number of required faulty ciphertexts and allowing for unknown single-byte faults.
However, note that a smaller number of required faulty ciphertexts is of minor importance in the persistent fault setting, as they ideally do not correspond to multiple physical fault injections.

\section{Persistent Fault Analysis}
\label{pf::sec::attack}
As our work mainly relates to \acf{PFA} as introduced by Zhang \etal in \cite{zhang:2018:ches} and \cite{zhang:2020:ches}, we introduce their notion and results in more detail.
A persistent fault \textit{persists} until the device is restarted, hence, likely for the duration of several executions.
Due to this behavior, \ac{PFA} is not restricted to a single, very precise fault injection in some last rounds of the computation, but typically faults persist and will occur in several rounds.
Hence, opposed to classical fault analysis, faults do not need to be injected during runtime.
Considering the implementation of a block cipher, a typical target for such a fault is memory holding for example a look-up table.

\ac{PFA} is a faulty-ciphertext-only analysis, hence, it is not required to repeat the encryption of a plaintext in both a faulty and fault-free setting such as in typical differential analyses. 
Instead, it relies on statistical analysis to recover the key.
Still, opposed to permanent faults or device defects which would allow for a similar analysis, the original behavior can be restored, i.e., the potential victim is oblivious to the attack once the device is restarted.

Given the first precise definition of \ac{PFA} in \cite{zhang:2018:ches}, Zhang \etal present the following fault model for persistent faults on block ciphers:

\begin{itemize}	
	\item Faults are injected prior to the encryption, altering a stored constant \eg in a look-up table
	\item The injected faults are persistent, hence, all following encryptions are subject to the faulty constant. To restore the original constant, the device needs to be restarted.
	\item Multiple corrupted ciphertexts can be collected.
\end{itemize}
	
Using this fault model, the authors show how \ac{PFA} can be used to attack the last round of a standard SPN block cipher when inducing a persistent fault into the Sbox.
For a successful recovery of the last round key, the adversary calculates the statistical distribution of each element of the ciphertext over a large amount of ciphertexts. 
Then, knowing which value $v$ of the Sbox has been altered, he can deduce the correct key element $k_j$ by $k_j = v \oplus t_{min}$, where $t_{min}$ is the value that has occured least in the ciphertexts. 
The authors present two further strategies to exploit persistent faults by either only knowing which Sbox element is faulty or both location and the value of the fault, and all three are then practically verified against an \ac{AES} implementation with a single persistent fault in its Sbox. 
Additionally, the authors show how to adopt their single-fault analysis to work in a multi-fault, \ie several faults per Sbox, setting.
Depending on the amount of faults, the authors show that the residual key entropy clearly increases, and suggest that for key entropies beyond brute-force search \ac{PFA} needs to be extended to the last but one round.

\enlargethispage{\baselineskip}

In \cite{zhang:2020:ches}, the authors practically perform \ac{PFA} using laser-based faults to attack an Sbox stored in the SRAM of a ATmega163L microcontroller.
Since knowing properties of the Sbox fault such as location and value is not feasible in this practical setting, they improve \ac{PFA} threefold:
By using the \ac{MLE}, the amount of ciphertexts needed to find the value for $t_{min}$ can be reduced.
Additionally, they present how to perform \ac{PFA} knowing neither the value of the fault nor its location, \textit{for single faults}. 
For multiple faults --- presented in the case of a double fault --- \ac{MLE} can be used to reduce the amount of ciphertexts, but location and value of the fault again need to be known.

\section{Statistical Persistent Fault Analysis}
\label{pf::sec::attack::ourwork}
So far, \ac{PFA} has been presented as a successful strategy to attack SPN based block ciphers such as AES. 
Although for single faults it was recently shown that \ac{PFA} is able to extract the secret key even in pratical settings \ie under a random fault, for multiple faults more details of the fault such as its value or location need to be known.

\subsection{Combining PFA and SFA}
Our work expands classical \ac{PFA} to show that \textit{even for multiple faults} neither the position of the fault nor the value of the faults needs to be known, allowing for faults beyond the control of the adversary.
To enable this, we combine \ac{SFA} with \ac{PFA} to \ac{SPFA}, gaining a fault model that offers the best of both worlds.
In particular, with respect to \ac{SFA}, we refer to the works of Fuhr \etal \cite{fuhr:2013:ftdc} and Dobraunig \etal \cite{dobraunig:2016:asiacrypt}.
Both works attack AES in the penultimate round, and use the statistical measure \acf{SEI} to identify the distance between the uniform distribution, which is expected in the error-free scenario, and the hypothetical distributions due to the injected fault.
For the correct key hypothesis however, the \ac{SEI} should peak as the faulty intermediate will cause some bias in the distribution.
Instead, wrong key guesses will not stand out because they would result in an approximately uniform distribution.
Note that attacking the last round of a typical \ac{SPN} is not beneficial for the attack, since the permutation layer only permutes number of occurrences for each (faulty) value, but all key hypotheses are equally likely.
Only if the error is distributed non-linearly \ie via a substitution layer, the correct key hypothesis stands out.
Hence, for a typical block cipher that is based on a \ac{SPN}, attacking the last but one round achieves the desired behavior.

  
\subsection{Fault Model}
\enlargethispage{\baselineskip}
Our relaxed fault model is as follows:
\begin{itemize}
	\item The adversary can insert faults prior to the encryption by making changes to stored constants in \eg a look-up table
	\item The impact of the fault can range from single-byte errors to a larger structure \eg several rows of a look-up table
	\item The injected faults are persistent, hence, all following encryptions are subject to the faulty constant. Original behavior can be restored via restart of the device.
	\item The exact position or value of the faults does not need to be known, the faults need to cause a non-uniform distribution.
	\item The adversary can collect multiple faulty ciphertexts.
\end{itemize}
Note that our attack does not require collecting fault-free ciphertexts or multiple fault injections and corresponding faulty encryptions to extract the entire key. 
Thus, the attack works in the permanent fault attack scenario as well.
In the following, we will explain how to conduct the attack step-by-step.

\subsection{Description of the Attack}
\label{pf::sec::ourwork::attack}

In accordance with the fault model, an adversary needs the following steps in order to perform the attack. 
In classical block ciphers, the Sbox look-up table offers a meaningful target for \ac{PFA}, hence, the following description targets the Sbox layer.
%
\subsubsection*{\textnormal{\textbf{Preparation}}}
First, the adversary needs to inject persistent faults into the Sbox look-up table prior to the encryption.
This faulty Sbox $\tilde{S}$ will be used throughout the encryption process and the faults will persist until the device under attack is restarted.
\revise{}{For our experiments, we consider a serial implementation of the Sbox \ie the same Sbox is used for all $16$ bytes of the state. 
However, \ac{SPFA} works similarly for parallelized implementations, e.g., featuring an individual Sbox for each byte of the state, since the attack operates on byte-level.}
\subsubsection*{\textnormal{\textbf{Collection}}}
Next, $N$ faulty ciphertexts $\tilde{c}$ are collected.
Note that due to the statistical nature of \ac{PFA}, neither plaintexts nor correct ciphertexts are needed for the analysis.
\subsubsection*{\textnormal{\textbf{Analysis}}}
Then, the adversary chooses for which target byte $t$ in which round $r$ he wants to observe the bias. 
Using an error-free implementation, the faulty ciphertexts are used to predict the value of said byte, under all possible key hypotheses, at position of the exploitable fault.
In other words, the decryption is computed backwards towards the fault.
To measure for which key hypothesis $\hat{k}$ the distribution of the target byte $t$ becomes non-uniform, the \ac{SEI} value $SEI(\hat{k})$ is then computed as: \enlargethispage{\baselineskip}
	$$ SEI(\hat{k}) = \sum_{\delta=0}^{2^{s}-1}\biggl({\frac{\#\{i\,|\, \tilde{S}_{r}^{(\hat{k}, \tilde{c}_i)}[t] = \delta\}}{N} - \frac{1}{2^{s}}}\biggr)^2 . $$ 

Here, $s$ is the bitlength of the target byte, hence, the $2^s$ gives the amount of different values $\delta$ the target byte have.
Since this hypothetical intermediate value $\delta$ is computed using a fault-free implementation, for a wrong key candidate, each value of $\delta$ is supposed to occur equally often using a sufficiently large amount of faulty ciphertexts.
However, for the correct key candidate, the distribution of the target byte becomes non-uniformly \ie some values of $\delta$ are more likely to occur.

\revise{}{Note that the SEI requires that $\tilde{S}_{r}^{(\hat{k}, \tilde{c}_i)}[t]$ does not linearly depend on the inputs.}
Also, the influence of faults that occur in rounds before (and, to some extent, after) the targeted round is negligible. This is due to the avalanche effect in typical block ciphers, spreading a respective error so that the distribution converges to a uniform distribution.
\enlargethispage{\baselineskip}
\subsubsection*{\textnormal{\textbf{Key Recovery}}}
The \ac{SEI} value is used as a ranking to identify the correct key, where the highest \ac{SEI} value indicates the best key candidate.
Depending on the target, in order to obtain the encryption key, additional steps such as reverting the key schedule might be needed.

\subsection{Comparison with classical PFA}

Comparing our model with classical \ac{PFA}, we want to highlight the advantages and disadvantages of \ac{SPFA}.

\subsubsection*{{Advantages}}
The adversary does not need to know the position of the fault even for multiple faults \ie unpredictable faults are allowed.
Although the authors of \cite{zhang:2018:ches} consider it an advantage that \ac{PFA} does not need any biased faults, with \ac{SPFA} the bias enables the adversary to exploit faults beyond his control, such as manipulating a bitstream to induce faults in the Sbox, as practically shown by Szwierczynski \etal in~\cite{pawel:2018:tc}.
As a direct consequence, multiple faults are possible and, to a certain extent, more faults even reduce the amount of faulty ciphertexts needed (cf. \Section{pf::sec::experiments::insights}).
Also, opposed to Zhang \etal, no further brute-force search is required as our method finds the corresponding key candidates instead of a key entropy.

\subsubsection*{{Disadvantages}}
For single faults in the Sbox, the \ac{SPFA} requires a larger amount of faulty ciphertexts.
Likewise, since the analysis requires computing the \ac{SEI} for all key guesses and all ciphertexts, the computation time is somewhat increased, especially when only a single byte is faulty.

\subsubsection*{{Countermeasures}}
When first introducing \ac{PFA}, Zhang  \etal examined its potency against countermeasures such as redundancy. 
Redundancy would require the same faulty Sbox in all instances. 
Due to the persistent nature of the fault, the same result will be computed if temporal redundancy is used. 
Spatial redundancy will of course detect the (persistent) fault if only one instance is faulted.
Also, \ac{SPFA} features the same fundamental assumption as \ac{PFA} with respect to attacker and fault model, hence, we assume a similar attack as presented by Pan \etal in \cite{pan:2019:date} to be possible using SPFA. 
Analyzing the possibilities when allowing multiple faults is an interesting idea for further research.

\subsection{Ineffectiveness Ratio of SPFA}
\label{pf::sec::ineffectiveness}
	Due to the persistent nature of the faults in \ac{SPFA}, we allow for a variety of faults to occur during the computations.
	Consequently, not all of these faults are exploitable by the underlying \ac{SFA}: 
	\ac{SFA} requires a faulty computation to occur at the target byte, typically in the penultimate round, but no further faults in the last round, to allow for the fault-free decryption to said target byte.
	Considering the amount of persistent faults in the SBox, for \ac{SPFA} to perform best we require that a) the target byte causes a faulty output after the SBox of the penultimate round and b) no additional fault that disturbs the decryption of the target byte occurs afterwards.
	
	To quantify this, we define the \textit{ineffectiveness ratio} as the amount of input values that cause fault-free outputs divided by all possible inputs.
	A small amount of faults results in a large ineffectiveness ratio, hence, a larger amount of ciphertexts is needed to achieve the desired bias in the penultimate round.
	At the same time, however, the probability for an error to occur in the last round is relatively small.
	Then again, a large amount of faults leads to a small ineffectiveness ratio, an erroneous computation at the target byte in the last but one round is very likely.
	However, the same holds for an error in the last round, hence, only a few ciphertexts will fulfill the requirements a) and b).
	
	Hence, there exists an optimal amount of error in the SBox such that \ac{SPFA} can exploit a maximum of the ciphertexts collected.
	The results of an experimental approach to find such said optimum are discussed in \Section{pf::sec::experiments::insights}.

	\subsection{Comparison with classical SFA}
	\label{{pf::sec::comparison_sfa}}
	Keeping in mind the discussion above, we would like to point out the strengths and limitations with respect to classical SFA.
	Classical \ac{SFA} typically relies on a transient fault model, i.e., the adversary needs to induce the fault during run-time, e.g., in the penultimate round. 
	Hence, an adversary is faced with problems such as timing as well as accessibility in order to launch the attack. 	
	While inducing the faults is more of a challenge, if successful, \ac{SFA} needs only a small number of faulty executions due to the determined injection of said faults.
	
	Opposed to this, due to its persistent nature, \ac{SPFA} allows for a more relaxed fault model, i.e., the adversary can induce faults offline.
	However, the adversary might not be able to influence the faults such that the desired faulty computations will take place 	
	at his target location, cf. \Section{pf::sec::ineffectiveness}.
	Hence, a larger number of faulty computations will be needed, which might be a limiting factor depending on the target.


%
%

%
%
%


\section{Practical Evaluation}
\label{pf::sec::experiments}

We implemented two case studies in C++ in order to simulate our attack.
Since computing the \ac{SEI} requires iterating over all key hypotheses and ciphertexts, its computation can be easily multi-threaded to reduce computation time.
In our setting, we used an IntelCore i7-6700 with 8 threads, but more threads could easily be used as well.

\subsection{General Set-Up}
\label{pf::sec::experiments::general}
The basic implementation of the attack is as follows, based on Sect.~\ref{pf::sec::ourwork::attack}: 
In our implementation, the Sbox is stored as a static table, and we manipulate (parts of) it to induce the persistent fault in every round (Preparation).
\revise{}{Note that for our experiments, we use a serial implementation with one SBox for all bytes. }
Then, we collect faulty ciphertexts and store them along with the encryption key (Collection).
Afterwards, we calculate the \ac{SEI} value for each key hypothesis, the key hypothesis with the highest \ac{SEI} value is our best key candidate (Analysis).
Due to the structure of the attack, the analysis phase can recover only a part of the key, e.g., 4 bytes in the case of AES, hence, it needs to be repeated to get all key candidates.
Note that we do not need additional physical faults for the remaining candidates.
Finally, we use the key candidates to recover the original (last round) key and then derive the original encryption key (Key Recovery).

\subsection{Case Study: LED}
\label{pf::sec::experiments::LED}

As a first case study for our analysis, we picked the LED block cipher \cite{poschmann:2011:ches}.

\subsubsection{Brief Overview of LED}
\label{pf::sec::background::led}
LED is an AES-like lightweight 64-bit block cipher first introduced at CHES 2011.
It comes with two standard key length of 64 resp. 128 bit and a block size of $64$ bit. 
In our attack, we will focus on the 64-bit variant consisting of $32$ rounds.

\begin{figure}[!htb]
	\centering
	\begin{subfigure}[b]{0.95\linewidth}
		\includegraphics[width=\linewidth]{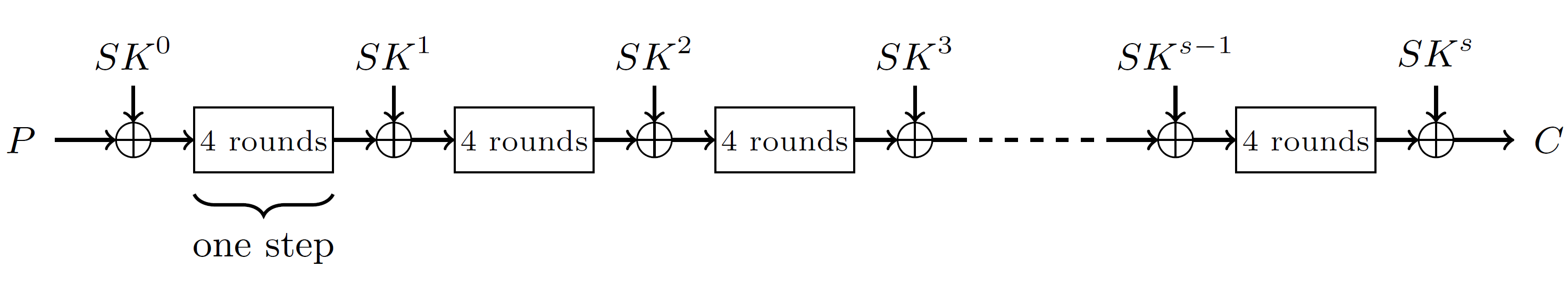}
		\caption{Workflow of the LED encryption.}
	\end{subfigure}
	\hfill
	\begin{subfigure}[b]{0.88\linewidth}
		\includegraphics[width=\linewidth]{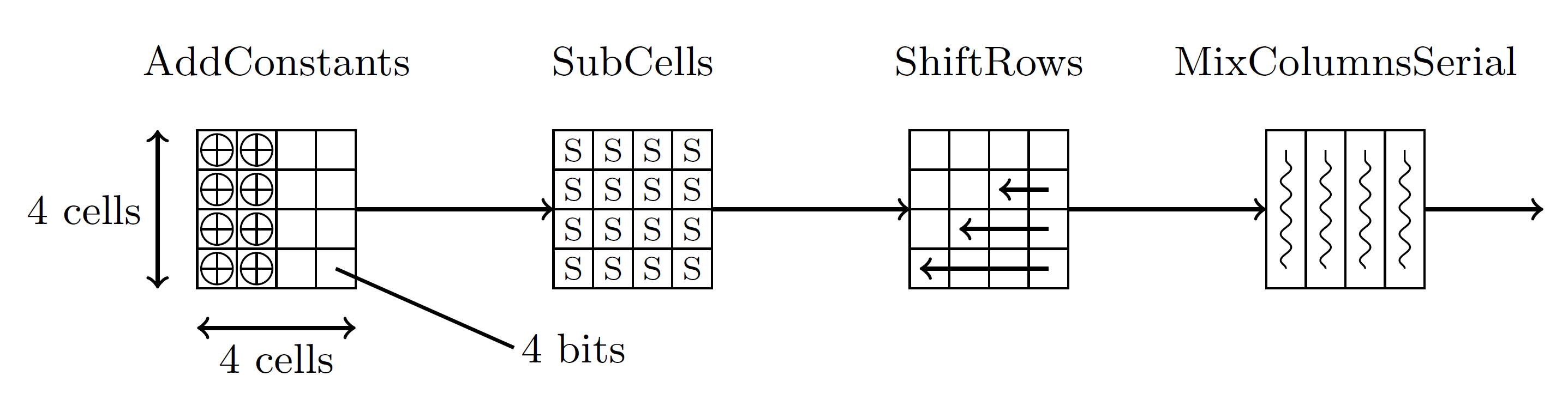}
		\caption{Round function.}
	\end{subfigure}
	\caption{Block diagram of LEDs round and encryption function as depicted in \cite{poschmann:2011:ches}.}
	\label{pf::fig::LED}
\end{figure}

%

\enlargethispage{\baselineskip}

The round function operates on a 4$\times$4-nibble state matrix similar to the AES state and features an AddConstants, a SubCells, a ShiftRows and a MixColumns operation, cf.~Fig.~\ref{pf::fig::LED}:
AddConstants (AC) combines the round constants with each nibble of the state via exclusive-or.
SubCells (SC) replaces each nibble of the state by its corresponding Sbox entry, here, LED makes use of the 4-bit Sbox of PRESENT. 
ShiftRows (SR) rotates each row of the state a fixed amount of positions to the left. 
MixColumns (MC) multiplies each column of the state with a matrix M and stores the result back in the same column.
Every four rounds, the AddRoundkey (AK) operation combines the key nibble-wise with the current state by exclusive-or.
To encrypt the 64-bit plaintext, it is loaded into the state matrix and likewise, after the final AddRoundKey, the ciphertext is extracted from the state matrix.

\subsubsection{Description of the Attack}
\label{pf::sec::experiments::led_attack}

\begin{figure}[!htb]
	\centering
	
	\includegraphics[width=\linewidth]{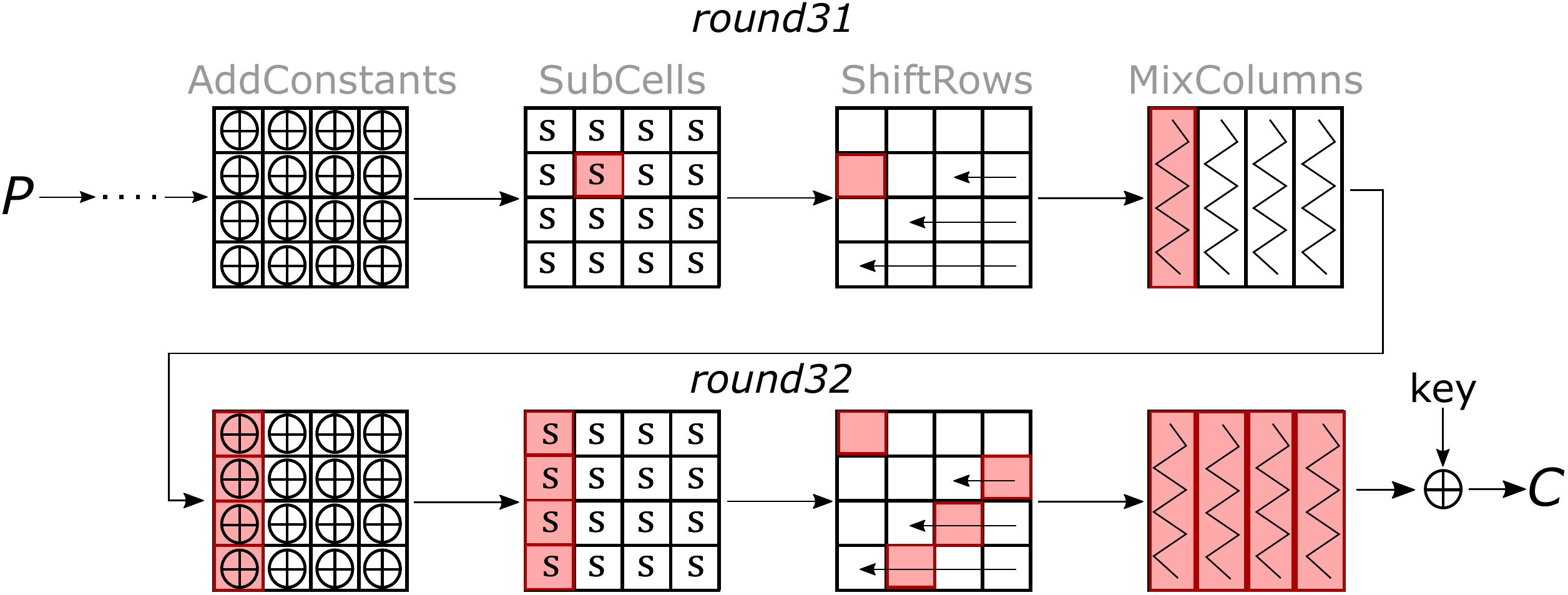}
	\caption{How a hitting a fault spreads in the last rounds for LED.}
	\label{pf::fig::LED_error}
\end{figure}

To induce the persistent fault for our experiments, we randomly exchanged one of the 16 entries of the Sbox, i.e., one entry is missing while another one is doubled. 
This Sbox is then stored and accessed in every round of the LED cipher, hence, every call to the Sbox could potentially result in a faulty return value.
In order to allow the error to create a bias in the distribution, we exploit an Sbox fault in round 31, cf. \Section{pf::sec::attack::ourwork}.
Figure~\ref{pf::fig::LED_error} depicts how said error gets spread during the following operations, until at the end of round 32 the entire state is faulty. 
Note that we only show a single fault, additional faults might happen simultaneously.
 
Our 4-nibble key hypothesis is XORed to one row of the ciphertext and then we compute backwards towards the Sbox in round 31, i.e., the red marked state entry in Fig.~\ref{pf::fig::LED_error}.
Opposed to AES, LED features a $MC$ operation in its last round.
Hence, to be able to revert the $MC$ operation in round 31 while taking into account the $MC$ operation in round 32, we need to analyse the state row-wise.
Computing the \ac{SEI}, we gather a ranking which key hypothesis is our best key guess and store this key candidate. 
These steps are repeated until we have a key guess for each row resp. nibble of the ciphertexts. 
After collecting key candidates for each row of the state, we need to revert the final $MC$ operation to recover the correct key.

\subsubsection{Results}
\label{pf::sec::experiments::led_results}
In order to gather a respectable statistic about its success, we conducted the attack on LED in an iterative manner for $50$ random key values, using random plaintexts for each encryption. 
Taking a look at the three phases of the attack, collection, analysis and key recovery, the analysis phase computing the \ac{SEI} requires most of the time. 
For each run, we collected $1000$ faulty ciphertexts. 
On our target machine, identifying the best key candidate, \ie computing the SEI for all key hypotheses, takes less than $90$ seconds per row.
Hence, completing the analysis phase for a single test case takes approximately $5$ minutes, whereas collection and key recovery take mere milliseconds. 
Considering the success rate of our analysis, for $50$ test cases \ie random keys and plaintexts, we were able to correctly recover all $50$ secret keys.


\subsection{Case Study: AES}
\label{pf::sec::experiments::AES}

LED's smaller state resp. key sizes drastically reduces the computation time for the SEI values, while its similarity to AES still allows for gathering first insights on the applicability on AES. 
Hence, after analyzing the potential of \ac{SPFA} on the LED cipher, we adapt the analysis to run on AES.
In comparison to LED, AES has a block size of $128$, hence, its state matrix is organized in 4$\times$4-bytes matrix.
With a key size of $128$ bit, an AES encryption takes $10$ rounds, hence fewer rounds than LED, and the MC operation is missing in the final round.
Instead of LED's AddConstants step, a different subkey is added at the end of every round of AES.
Similarly to LED, we exchange part of the AES Sbox with random entries, starting with only one of the $256$ entries.
Note that the amount of faults does not influence the course of the analysis, hence, the following description is generic.
In \Section{pf::sec::experiments::insights}, we present a more detailed overview on how the different fault scenarios influence the analysis.

Following Fuhr \etal \cite{fuhr:2013:ftdc}, we attack the penultimate round of AES, cf. \Section{pf::sec::attack::ourwork}.
The attack can recover a quarter, i.e., four byte, of the key, hence, to retrieve the whole key, it needs to be repeated four times.
Naturally, each run utilizes a new column of the faulty ciphertexts to regain the corresponding part of the key. 

\begin{figure}[!htb]
	\centering
	
	\includegraphics[width=\linewidth]{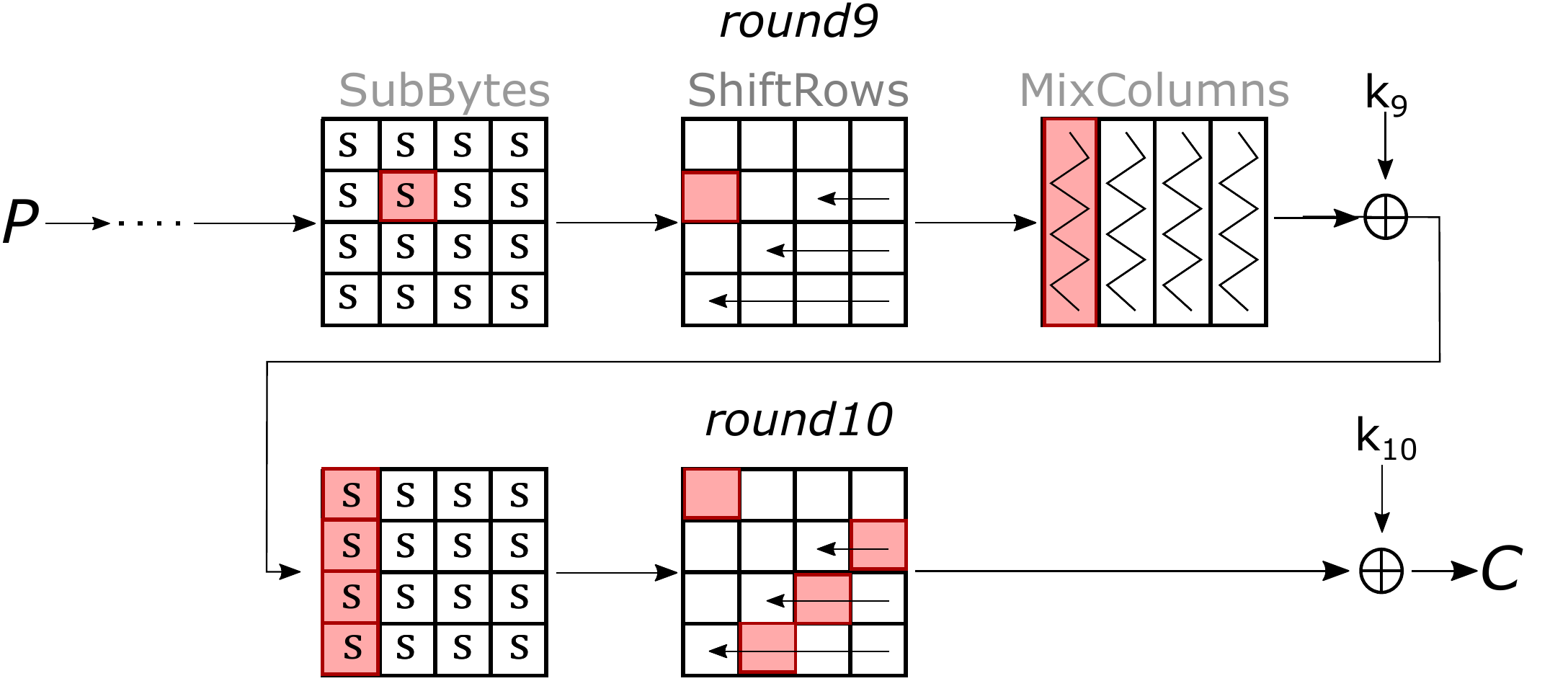}
	\caption{How a fault spreads over the last two rounds of AES.}
	\label{pf::fig::AES_error}
\end{figure}

Figure~\ref{pf::fig::AES_error} depicts how a fault in round 9 further influences the state.
Compared to LED, it is clearly visible that the missing $MC$ operation in the final round causes the error to influence fewer bytes in the final state.

\enlargethispage{\baselineskip}

In a first step, we use the 4-byte key hypothesis $k_{10}$ and one column of the faulty ciphertext $\tilde{c_{i}}$ with $i=0,1,2,3$ as input to invert the final $AK$ layer in round 10, then, we calculate backwards to the $SB$ operation in round 9. 
Note that Fuhr \etal showed that using the \ac{SEI} distinguisher, instead of inverting all operations, it suffices to merely invert the last round and the $MC$ operation of round 9, omitting the $AK$ layer in round 9 completely. 
Hence, the hypothetical state $\tilde{S_9}$ after the Sbox fault in round 9 can be computed by $\tilde{S_9} = MC^{-1} \circ SB^{-1} \circ SR^{-1}(k_{10} \oplus\tilde{c_{i}})$, cf. \cite{fuhr:2013:ftdc}.
For our target byte in $\tilde{S_9}$, i.e., the red state in $SB$ of round 9 in Fig.~\ref{pf::fig::AES_error}, the \ac{SEI} is then computed for every key hypothesis.
Again, the key hypothesis with the highest \ac{SEI} value is our best key guess.

Naturally, we expect a higher computation time due to key and state size compared to LED.
For each column of the AES state, the complexity is $2^{32} \cdot SEI(\hat{k})$.
Depending on the amount of ciphertexts needed, a naive implementation will take several hours. 
Hence, to enable a larger study on the effect of fault position, amount of faults, and number of ciphertexts needed, we fix $2$ byte of the key.
The corresponding results can be found in the next section. 
\revise{}{Using the insights of said study, we confirmed our results by conducting a full-key search, needing only a negligible overhead of ciphertexts.}


\enlargethispage{\baselineskip}
\subsection{Insights on AES}
\label{pf::sec::experiments::insights}
\begin{figure}[tb]
	\centering
	\begin{subfigure}[b]{0.48\linewidth}		
		\includegraphics[width=\linewidth]{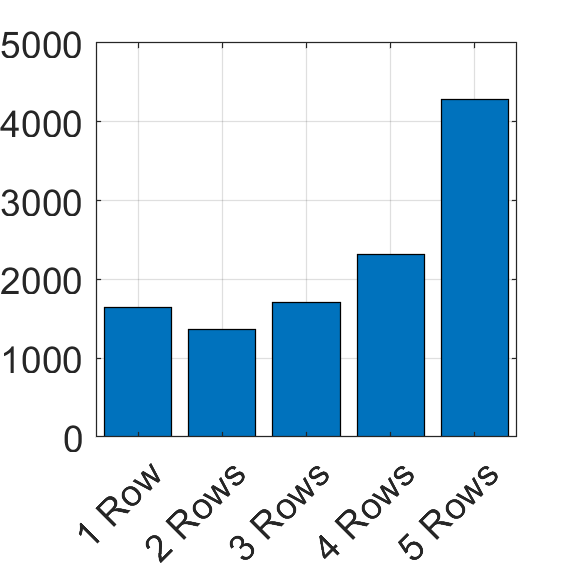}
		\caption{Less than 5,000 faulty ciphertexts needed.}
	\end{subfigure}
	\hfill
	\begin{subfigure}[b]{0.48\linewidth}
		\includegraphics[width=\linewidth]{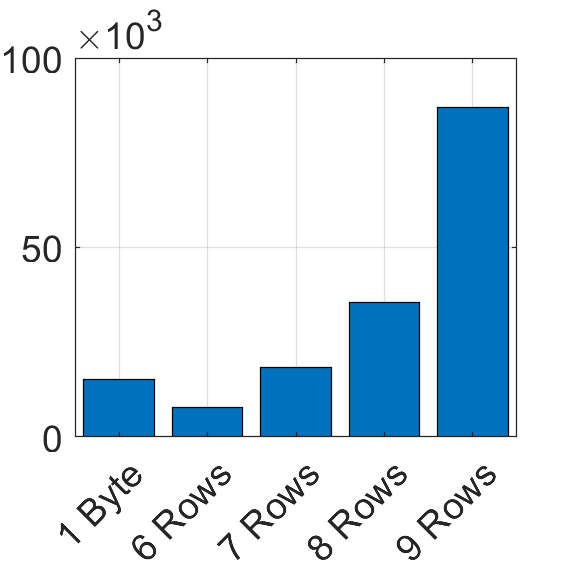}
		\caption{More than 10,000 faulty ciphertexts needed.}
	\end{subfigure}
	\caption{Experimental mean of amount of faulty ciphertexts needed depending on error, computed with 2 keybytes fixed.}
	\label{pf::fig::experiments::amountoffaults}
\end{figure}

We conducted different fault scenarios on AES to gain insights about how they influence the success of our attack. 
First, we only exchanged one entry of the $256$-byte Sbox. 
Since in this scenario we expect an exploitable error to occur only every $64$ ciphertexts, the amount of faulty ciphertexts needed is higher than for the single-byte fault in the LED Sbox.
Our initial estimation suggests that an amount of $30000$ ciphertexts allows for a sufficient amount of faulty ciphertexts to conduct our analysis. 
In order to specify this assumption, we ran $10$ tests with increasing amounts of random plaintexts resp. faulty ciphertexts for each column.
Our analysis shows that the amount of faulty ciphertexts strongly depends on the input.
For a single byte error, the amount ranges from 7500 to 23000 ciphertexts (Fig~\ref{pf::fig::experiments::amountoffaults}).

Next, we want to investigate how ``faulty'' the Sbox can be, \ie how many faulty entries still lead to a successful attack.
Hence, we successively exchanged the rows of the Sbox with rows with faulty entries and launched the attack with an increasing number of randomly chosen ciphertexts.
See \Section{pf::sec::experiments::logic_friday} for a different approach to induce faults into the Sbox.

\enlargethispage{\baselineskip}

Starting with a single faulty row, the correct key could be found surprisingly fast compared to the single byte case \ie on average 1645 ciphertexts suffice.
We continued to exchanged row after row until half of the Sbox was corrupted.
Figure~\ref{pf::fig::experiments::amountoffaults} aggregates our results.
Our experiments show that the optimal amount of faults with respect to computational complexity is around two rows \ie $\frac{1}{8}$ of the Sbox.
Further experiments have shown that the optimum for a serial implementation resp. only faulty SBoxes is $32$ \ie two rows, confirming our results. For parallelized implementations that instantiate an Sbox for each byte, the amount of faulty entries with the optimal ineffectiveness ratio also depends on the overall number of faulty Sboxes, cf. \Section{pf::sec::ineffectiveness}.



\begin{figure}[!t]
	\centering
	\begin{subfigure}[b]{0.48\linewidth}
		\includegraphics[width=\linewidth]{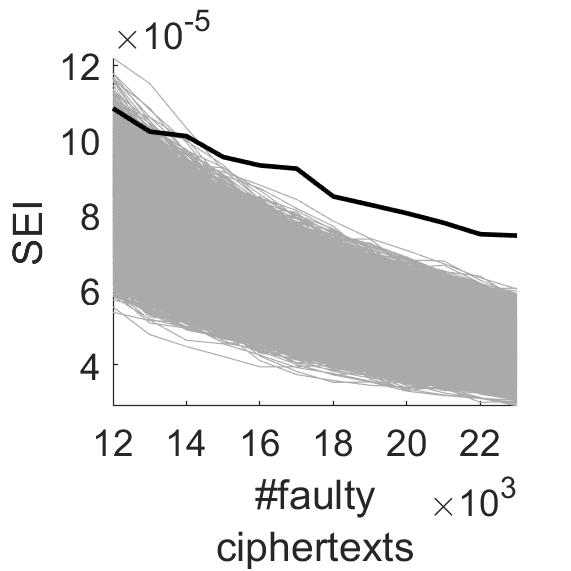}
		\caption{Result for a single faulty byte, on average.}
	\end{subfigure}
	\hfill
	\begin{subfigure}[b]{0.48\linewidth}
		\includegraphics[width=\linewidth]{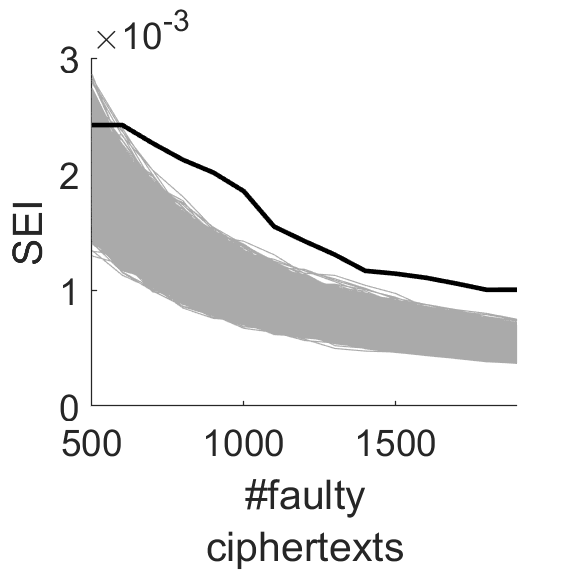}
		\caption{Result for two faulty rows, best case.}
	\end{subfigure}
	\caption{Number of ciphertexts for a successful attack, experimental results.}
	\label{pf::fig::numberofct}
\end{figure}


\enlargethispage{\baselineskip}
\begin{table}[hbt]
	\centering
	\begin{tabular}{c||c|c|c||c|c}
		\multirow{2}{*}{\#faults}               & \multicolumn{3}{c}{\#ciphertexts}  & \multicolumn{2}{c}{complexity} \\ 
		& w/o MLE & MLE & our work & Zhang & our work \\
		1 & 2273 & 1641 & 15650 & 0 & $2^{50}$ \\
		2 & ca.2000 & n/a & 7775 & $2^{16}$ & $2^{50}$ \\ 
		8 & ca.2000 & n/a & 2008 & $2^{50}$ & $2^{50}$ \\ 
		16 & ca.2000 & n/a & 1643 & $2^{64}$ & $2^{50}$ \\ 
		
	\end{tabular}%
	\caption{Comparison of the complexity of our experimental results with the works of Zhang et. al.}
	\label{pf::tab::comparison}
\end{table}


Comparing our results with \ac{PFA} as presented so far, we experience a trade-off between the amount of ciphertexts needed for the attack, and the residual key entropy resp. complexity as shown in \Table{pf::tab::comparison}:
For up to $16$ faults, \ie one row of the Sbox table, Zhang \etal presented results for \ac{PFA} without using \ac{MLE} in \cite{zhang:2018:ches}.
Although our analysis is computationally expensive \ie $2^{50}$ for the complete key, so is the required brute-force search to find the remaining key bits.
For up to eight faults, less than $50$ bits need to be brute-forced using classical \ac{PFA}, with exactly eight faults, both \ac{SPFA} and \ac{PFA} perform equally, and for more than eight faults \ac{SPFA} performs better.

Although the amount of faulty ciphertexts is not of great importance in the persistent setting, our experiments show that for eight faults, the average amount of ciphertexts needed for \ac{SPFA} is equal to \ac{PFA} without \ac{MLE}. 
Using \ac{MLE} reduces the amount of ciphertexts needed for \ac{PFA}.
The authors do not extent their results to more than two faults, but similarly to the results in \cite{zhang:2018:ches}, we expect this amount to remain constant in the multi-fault setting.
For $16$ faults, the average amount of ciphertexts needed in our experiments equals the amount for \ac{PFA} with \ac{MLE}.
For more than $16$ up to $24$ faults, this amount is further reduced using \ac{SPFA}, hence undercutting the expected number needed, even when using \ac{MLE}.

Summing up, for more than eight faults, \ac{SPFA} outperforms \ac{PFA} in terms of complexity and number of faulty ciphertexts needed.
Our experiments show that \ac{SPFA} has its optimal configuration around $32$ faults, hence, should be the method of choice for larger numbers of faults.
%

\enlargethispage{\baselineskip}

\subsection{Cutting Wires of Sboxes}
\label{pf::sec::experiments::logic_friday}
The results presented so far mostly correspond to the faulted Sbox-memory scenario, e.g., for microcontroller implementations. 
In this scenario, the adversary has control about the number of faults to inject.
Now, we discuss the hardware or faulty bitstream case:
Here, an adversary might not be able to manipulate the Sbox that freely. 
Instead, especially in the multi-fault setting, errors could occur by destroying parts of the Sbox storage structure \eg by cutting a wire or manipulating bits of a bitstream as presented in \cite{pawel:2018:tc}.
Especially due to recent successful attacks against Xilinx bitstream encryption \cite{maik:2020:usenix}, using \ac{SPFA} when meddling with the bitstream to create persistent errors becomes an interesting target.
Such an error would cause faults at random positions in the Sboxes, and not necessarily on byte level, but instead only single bits of certain entries might flip.
To simulate such faults, we used the free software tool ``Logic Friday''\footnote{Logic Friday can be found at \url{https://download.cnet.com/Logic-Friday/3000-20415_4-75848245.html}}.
Logic Friday takes as input either a truth table, an equation or a gate diagram and enables the user to modify either of them and views the resulting other two.
Hence, entering its truth table, it is possible to manipulate the gate diagram of the Sbox by removing wires or exchanging constant. 
The result is a new truth table which can again be used in our analysis.

\enlargethispage{\baselineskip}
\subsubsection{AES}
\begin{figure}[!t]
	\centering
	\begin{subfigure}[b]{0.88\linewidth}
		\includegraphics[width=\linewidth]{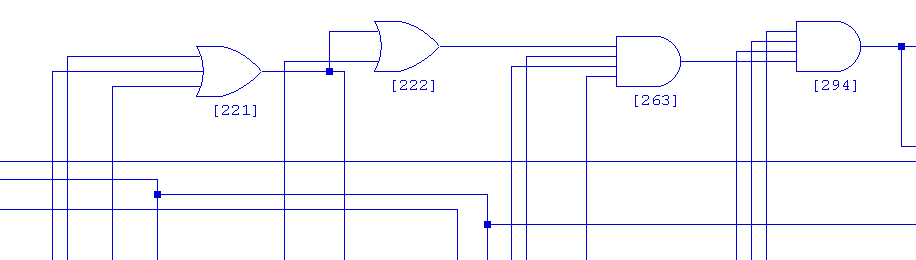}
		\caption{Original wiring at gate 222.}
	\end{subfigure}
	\newline
	\begin{subfigure}[b]{0.88\linewidth}
		\includegraphics[width=\linewidth]{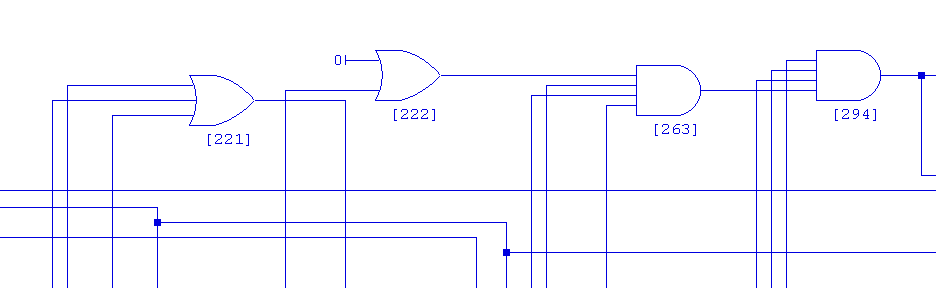}
		\caption{First input of gate 222 fixed to 0.}
	\end{subfigure}
	\caption{Modifying wires of the AES Sbox}
	\label{pf::fig::logic_friday}
\end{figure}

In order to induce a more natural random fault into the AES Sbox, we mapped the Sbox truth table to a corresponding gate diagram.
Then, in the resulting gate diagram, a random gate is picked and modified, \eg an input is fixed to a constant.
\Figure{pf::fig::logic_friday} depicts gate222 before and after fixing one of its inputs to a constant $0$.

\begin{figure}[tb]
	\centering
	\begin{subfigure}[b]{0.48\linewidth}
		\includegraphics[width=\linewidth]{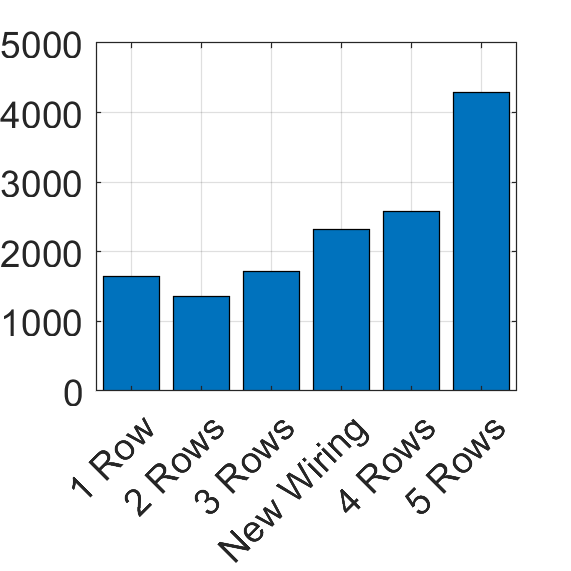}
		\caption{New wiring in comparison with faulty rows.}
	\end{subfigure}
	\hfill
	\begin{subfigure}[b]{0.48\linewidth}
		\includegraphics[width=\linewidth]{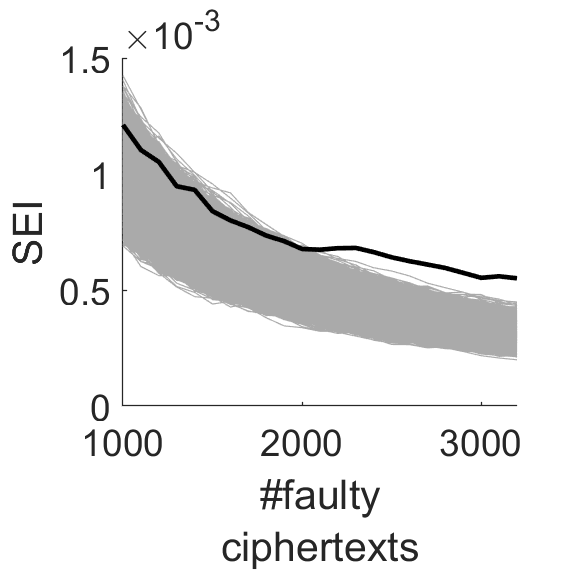}
		\caption{Correct key on average after approx. 2200 faulty ciphertexts.}
	\end{subfigure}
	\caption{Results of our analysis with the Sbox altered.}
	\label{pf::fig::logic_friday_result}
\end{figure}

The faults in the resulting Sbox spread much wider than in our original test cases, \ie there are faults in almost every row but some of them are only faults on bit level.
In total, a fault is introduced in approximately a fourth of the Sbox.
As expected, a successful analysis is possible with this Sbox configuration.
Again, the amount of needed faulty ciphertexts strongly depends on the inputs. 
For each column, we ran $10$ tests on different plaintexts for our tested key.
The results are similar to those of the Sbox configuration with four faulty rows, as both results in a fourth of the Sbox being corrupted.
Hence, for our analysis, it does not matter if the fault is spread over the Sbox or not, only the amount of faulty values influences how many faulty ciphertexts are needed, as shown in Fig.~\ref{pf::fig::logic_friday_result}.
Also, in this setting, the faults are often only on bit level, \ie only a single bit is flipped, whereas in Sect.~\ref{pf::fig::experiments::amountoffaults}, the faults are usually on byte level.
Still, as the number of faulty elements is the same, the nature of the fault per element does not matter for the complexity of the attack.
Figure~\ref{pf::fig::logic_friday_result} summarizes these results.

\enlargethispage{\baselineskip}

%

\enlargethispage{\baselineskip}
\section{Conclusions}
\label{pf::sec::conclusion}
We combine\revise{}{the well-known concept of SFA by introducing persistent faults as used in PFA} to allow for a more relaxed fault model.
Opposed to \ac{PFA} presented so far, even for multiple faults, the adversary needs to know neither the value nor the position of the fault. 
Hence, additional attack scenarios, such as random manipulations of the SBox, become feasible.
As exploiting multiple faults is feasible and even beneficial for the analysis, \ac{SPFA} allows for random faults in larger settings \eg faults on gate level or manipulation of bitstreams.
We verified our results by simulating our attack for the block ciphers LED and AES. 
For AES, we additionally identified the optimal number of faults with respect to computational complexity.
Our experiments show that manipulating almost half of the Sbox still allows for a successful key recovery, albeit needing a larger amount of faulty ciphertext and computation time, respectively.
For future work, it might be interesting to analyze if exploiting ineffective faults such as in introduced in Statistical Ineffective Fault Analysis \cite{dobraunig:2018:ches} optimizes our attack even further:
\revise{}{These ineffective faults lead to seemingly fault-free ciphertexts, undetectable by most countermeasures, while still showing an exploitable non-uniform distribution. Also, it might be an interesting idea to test \ac{SPFA} against other Sbox designs such as e.g. the Canright Sbox for AES.}



\section*{Acknowledgments}
We thank our reviewers for their valuable input and additional results to our paper. 
This work was supported in part by grant ERC 695022, NSF CNS-1563829, and BMBF 16KIS0602.
\enlargethispage*{\baselineskip}

\bibliographystyle{alpha_short}
	
\bibliography{localbib, bibliography}

\newpage




\end{document}